\documentclass[showpacs,preprintnumbers,amsmath,amssymb]{revtex4}
\usepackage{graphicx}
\usepackage{dcolumn}
\usepackage{amssymb}

\begin{document}

\preprint{APS/123-QED}

\title{ Temperature Dependence of Fluctuation Time Scales in Spin Glasses }

\author{G.~G.~Kenning$^{\ddagger}$}

\author{J. Bowen$^{\dagger}$}
\author{P. Sibani$^{\ddagger\ddagger}$}
\author{G.~F.~Rodriguez$^{\dagger}$}
\affiliation{$^{\dagger}$Department of Physics, Indiana University of Pennsylvania\\
Indiana, Pennsylvania 15705-1098}
\affiliation{$^{\ddagger}$Department of Physics, University of California\\
Riverside, California 92521-0101}
\affiliation{$^{\ddagger}$Institut for Fysik og Kemi, SDU DK5230 Odense M. Denmark\\
Odense, Denmark }

\setlength{\baselineskip}{20pt}

\date{\today}

\begin{abstract}
\setlength{\baselineskip}{20pt} 
Using a series of fast cooling protocols we have probed 
aging effects in the spin glass state as a function of 
temperature. Analyzing the logarithmic decay found at 
very long time scales within a simple phenomenological 
barrier model, leads to the extraction of the fluctuation
 time scale of the system at a particular temperature.
  This is the smallest dynamical time-scale, defining a 
  lower-cut off in a hierarchical description of the 
  dynamics.  We find that this fluctuation time scale, 
  which is approximately equal to atomic spin fluctuation
   time scales near the transition temperature, follows a
    generalized Arrhenius law. We discuss the hypothesis
     that, upon cooling to a measuring temperature within
the spin glass state, there is a range of dynamically 
in-equivalent configurations in which the system can 
be trapped, and check within a numerical barrier model
 simulation, that this  leads to sub-aging behavior 
 in scaling aged TRM decay curves, as recently discussed 
 theoretically~\cite{Sibani09}.

\end{abstract}

\maketitle


Systems far from equilibrium display a range of interesting properties that are similar for what appear to be very different situations.  Spin glasses are prototypical non-equilibrium systems that provide an interesting and accessible vehicle for the investigation of non-equilibrium statistical mechanics.  Among the most striking properties of spin glasses are the time-dependent dynamics.  Measuring time-dependent variations of the magnetization, as a function of temperature, within the spin glass state, allows a probe of accessible phase space which is characterized by highly degenerate, but distant, free energy minima.  In previous studies, many similarities have been found for measurements over a wide range of temperatures.  For example, aging occurs at all temperatures below $T_g$, the spin glass transition temperature\cite{Ocio85, Alba86, Alba87}.  The aged data, at different temperatures, also have some differences which, although subtle, require further study.  These non-equilibrium properties are directly related to the underlying phase- and real-space structure of the system. 

The dynamics of complex systems are known to depend strongly on the initial
conditions\cite{Zov02,Rod03,Parker06}, defined
as the state of the sample at the time the experiment begins.
i.e. $t=0$s at the beginning of the isothermal aging process, or alternatively at the end of the quench. For spin glasses, the initial conditions depend on the thermal history occurring in the relatively short time period starting when  the sample temperature crosses 
$T_g$, and ending when  the measurement temperature $T_m$ is reached.  We report in this paper aging measurements  using the thermoremanent magnetization (TRM) decay for a range of measuring temperatures below $T_g$.  The decay is measured at times $t > t_{\rm w}$, where $t_{\rm w}$ is the waiting time, defined as the interval between the time when the sample reaches the measuring temperature,
$T_{m}$, and the time when the magnetic field, in which the sample is cooled through the transition temperature, is cut to zero.
The observation time $t_{\rm obs} = t -t_{\rm w}$ is the time during which the data are taken. This quantity is traditionally called $t$ and is used an independent time variable for TRM measurements.

In a previous paper\cite{Ken06} we found, using a fast-quench cooling protocol with 'zero additional waiting time' (ZTRM),  that the ZTRM decay showed a fairly standard TRM-like decay with a small effective waiting time ($\approx$ 19s) which was associated with the cooling protocol. After several thousand seconds however, the ZTRM became logarithmic in time out to $10^5$s, the longest time measured. 
Repeating the same procedure, but adding a short waiting time after the zero waiting time" protocol, the $t_{\rm w}$ dependent part of the decay ended at a time significantly longer than that found from the zero waiting time protocol, but nevertheless ultimately decayed into the very same logarithmic time dependence. Subtraction of this logarithmic term from the long waiting time TRM decays (where scaling holds) utterly destroys the ability to scale the decay curves, implying that the logarithmic decay is not an additive term. These two empirical findings suggest that the states associated with the logarithmic decay are not an independent contribution to the magnetization decay but are intrinsic to the aging curves themselves. It was also shown that a hierarchical Barrier Model simulation with a uniform initial state distribution produced a series of TRM decays (for different waiting times) that decayed into a common waiting time independent logarithmic decay. The same conclusion was reached in a theoretical model~\cite{Sibani09} where the disappearance of the $t_{\rm w}$ dependence of the TRM decay and sub-aging behavior result from a spatial heterogeneity of the initial configuration. The concept of spatial heterogeneity with a distribution of time scales is not new. Chamberlain~\cite{Chamberlain99} has observed nonresonant spectral hole burning in a AuFe($5\%$) spin glass sample. He found that the decay observed after applying large-amplitude low-frequency magnetic fields implies a distribution of relaxation times corresponding to a heterogeneous system composed of many domains, each with its own characteristic temporal behavior. Montanari and Ricci-Tersenghi~\cite{Montanari03} have applied microscopic Fluctuation Dissipation Relations to a given disorder realization of a highly frustrated ferromagnetic Ising Model. They find that single spins can exhibit heterogeneity of time scales. In this paper we apply the concept of heterogeneity to the Barrier Model of spin glasses through the imposition of an initial distribution of states with each state corresponding to an independent domain.

Numerical studies of multi-valley energy landscapes of complex model systems, including Edward-Anderson spin glasses\cite{Sibani93,Sibani94,Schön00}, have charted out the ``shape" of landscape valleys, i.e. regions of phase-space enclosing local energy minima. Defining for convenience the energy of a specific local minimum to be zero, the number of microscopic configurations ${\cal D}(E)$ within an interval of width $dE$ around energy $E$ was evaluated by exhaustive enumeration and other techniques. In most cases, this so-called local density of states has, for a range of $E$ values, a near exponential growth as a function of $E$, i.e. ${\cal D}(E) \approx c\exp(E/E_0)$, where $c$ and $E_0$ are constants. The parameter $E_0$ has the role of a local `~transition temperature": for $T>E_0$, the valley in question is not thermally accessible since the large number of configurations near its rim creates a ``top heavy" Boltzmann quasi-equilibrium distribution. In this regime, the thermalization process roams freely through the states near the rim of the valley. When $T<E_0$, the valley begins to act as a trap, as the configurations of low energy then carry a large probability. Interestingly, the values found numerically for $E_0$ for the Edwards-Anderson spin-glass are close to the critical temperature of the model~\cite{Sibani94, Sibani99}. The numerical results just described were obtained for models of relative small size, and are applicable to the localized and bounded regions of a spin-glass which are in a state of local thermal equilibrium, i.e. thermalized domains well known from numerical investigations\cite{Kisker96}.    Assuming that spatially bounded regions with a near exponential local density of state exist in a spin glass, a dynamical   ``phase transition" occurs as the temperature decreases below the transition temperature, whereby a hierarchically structured barrier space emerges for the volume.  At the end of the quench, and the beginning of the 
isothermal aging process,
 several domains will be formed, each characterized by a maximum barrier. The 
system as whole is
 then characterized by an initial barrier distribution, which underlies the 
spatial heterogeneity of the system. Within each domain, the current configuration will be trapped in the hierarchy at a position mirroring the random configuration of the system right above the transition, and hence a position which appears   random in relation to the hierarchical barrier structure.  It is plausible that this process repeats itself as the temperature is lowered leading to a continuum of phase transitions as the temperature is lowered. The free energy barriers within the space are strongly temperature dependent increasing rapidly in size as the temperature is lowered in the spin glass phase. As the temperature decreases large barriers grow very large, effectively making regions of phase space inaccessible to each other while smaller barriers grow and replace larger barriers maintaining the hierarchical structure at each temperature. Applying the relationship between barrier height and temperature from Lederman 
et al.\cite{Led91} to the large
barrier, associated with the time at which the TRM vanishes in isothermal aging 
at $T_m$, we see 
that a small temperature decrease from   $T_m + \Delta T$ to $T$ ($\Delta$T $\approx$ 300mK in $Cu_{.94}Mn_{.06}$ the sample used in this study)  brings a 
barrier of this large size below  the barrier associated with the fluctuation time scale at $T_m$.  Hence, a portion 
of phase space which 
 essentially looks flat at $T_m + \Delta T$ develops a full hierarchy of 
barriers when cooled to $T_m$. As the temperature decreases from $T_m + \Delta T$ to  $T$, the state gets trapped in some valley of the barrier space and can  migrate out of this portion of the space only if the surrounding barriers never grow larger than the barriers associated with the cooling time scale. As the measuring temperature is approached the barrier space at that particular temperature unfolds and within a single particle picture the system enters and occupies a single state within the barrier space. There is no apriori reason for the occupation of any particular initial state other then the constraints due to growing barriers presented in the above arguments.  In a large system composed of many volume limited independent domains,  the initial state occupations of each domain, would on average be random producing  a uniform distribution leading to a logarithmic decay for $t >> t_{\rm w}$\cite{Ken06,Sibani09}. If the fluctuation time scale of the system is small and the time for hopping over the largest adjacent barrier associated with that particular state (i.e. the local barriers confining the state) is much less than the waiting time then full aging follows. If time for hopping over the smallest adjacent barrier (i.e. the local barriers confining the state) is much greater than the waiting time then the state is effectively frozen  and no aging would be observed. These states would however decay for measuring times $t >> t_{\rm w}$.  

In the first section, we report TRM measurements made at a series of temperatures below $T_g$.  In the second section we reanalyze the temperature dependent ZTRM decays obtained\cite{Ken06} using rapid cooling protocols at a series of temperatures below the spin glass transition temperature  $T_g$. Finally we use the experimental results of the first two sections to determine parameters used to produce a series of  Barrier Model simulations of the temperature dependent TRM decays. We then compare the scaling behavior of the experimental and simulation data.

\section{Temperature Dependent TRM Data}
\label{sec:SclTemp}

In a previous paper\cite{Rod03} the results of various temperature quenching protocols were reported at T = 26K or $T_{r} = 0.83T_{g}$ for $Cu_{.94}Mn_{.06}$. We now examine other measurement temperatures. Previous studies\cite{Ocio85, Alba86, Alba87} using conventional cooling techniques, have found the  scaling parameter $\mu$(determined by scaling the data with an effective waiting time $t_{\rm w}^{\mu}$) to be relatively constant (approximately $\mu$=0.9) over thetemperature range  $.4T_g-.9T_g$. Outside of this range $\mu$ decreases.  We have repeated our decay experiments (using similar fast cooling protocols) at the same waiting times as before, but now at different temperatures in order to examine the temperature dependent behavior under more controlled conditions.  

Starting from the same high temperature, $T_{h}$ =
35K, the sample is then rapidly cooled through the transition temperature $T_g$ = 31.5K to a measuring temperature $T_{m}$.  At any particular measuring temperature a full range of TRM experiments were performed with waiting times ranging from 50s to 10,000s. Four different measuring temperatures were explored: $T_{m}$ = 12.6K, 18.9K, 28.9K and
29.9K, corresponding to reduced temperatures units $T_{r}=T_{m}/T_g$ = 0.4, 0.6, 0.9, 0.95
respectively. The effective cooling times for the new temperatures
are similar to the fast cooling protocol employed for the .83 $T_g$ study (i.e.  $t_{c}^{eff} \approx 20s$).  The cooling rate was however increased from 1K/s to 2K/s for the lower temperatures.

Figure 1 shows the thermoremanent magnetization decay curves for different temperatures spanning the range of temperatures we probed.  It can be observed, as has previously been observed\cite{Alba86}, that there are qualitative similarities for aging data at different temperatures.  There are however several significant differences.  To begin with, the width of the decays (over the entire span of waiting times) for the two sets of data on the outer temperature ranges, .95 $T_g$ and .4 $T_g$, is slightly smaller than the width of the decays for the temperatures in the intermediate region. For all of the decays, at a particular measuring temperature, the magnetization values at the observation time of .5s are similar but show a small increase as the waiting time increases.  The curves then separate and at very long times appear as if they are going to come back together.  In a previous paper\cite{Ken06} we showed that the curves actually recombine, within experimental observation times, for very short waiting times ($t_{\rm w} \leq 45s$).  For the waiting times used in this study, the recombination of the long waiting time curves is clearly outside of experimental measuring times.  A second and more obvious difference is the magnitude of the initial remenance for each temperature. Assuming similar behavior for all of the temperatures the increased remanence suggests that, as the temperature is lowered, the logarithmic term  would extend over a much longer timescale and make up a significantly larger proportion of the total remanence.  For the high temperature data the aged term is the majority of the remanence.

\begin{figure} 
\resizebox{4.0in}{3.0in}{\includegraphics{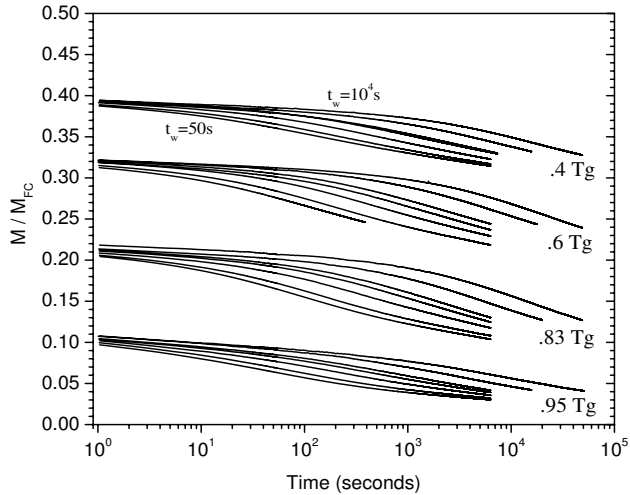}}
\caption{\label{fig:bob1}Experimental TRM curves of $Cu_{.94}Mn_{.06}$ using fast cooling protocols at four different temperatures. Each set of decay curves at given temperature display TRM decays for waiting times of tw= 50s, 100s, 300s, 630s, 1000s, 3600s and 10000s. In each series the 50s curve is the bottom curve and the magnetization increases systematically for increasing waiting time to the uppermost 10000s curve. The time on the x-axis begins after the magnetic field is cutoff.}
\end{figure}

\section{Analysis of ZTRM Curves}
\label{sec:SclTemp2}
We start by reanalyzing some  previously published data\cite{Ken06}.  Figure 2 is a plot of ZTRM decays measured out to $10^5$ seconds for four temperatures, and out to 1$0^3$ seconds for one temperature, i.e.  T= .6 $T_g$.   This data shows a long time logarithmic tail from approximately 6x$10^3$ seconds out to 1x$10^5$ seconds for each of the temperatures (excluding $.6T_g$).  This tail appears to begin after the $t_{\rm w}$ dependence of the cooling protocol comes to an end. Within the Barrier Model\cite{Ken06} we observe this behavior of aging decay giving way to a logarithmic decay when a uniform distribution is utilized.  The extrapolation of the logarithmic term to zero magnetization is an intriguing result.  The uniform distribution within the Barrier Model suggests that aging can occur with waiting times up to approximately a maximum time, $\tau_{\rm max}$, which is associated with the maximum barrier set up during the cooling process.  If the waiting time is significantly less than the maximum time then the aging curves will eventually decay into a common logarithmic decay.  This result implies that even though time scaling may occur in the usual aging regime, time scaling eventually breaks down as the logarithmic decay is entered.

\begin{figure}
\resizebox{4.0in}{3.0in}{\includegraphics{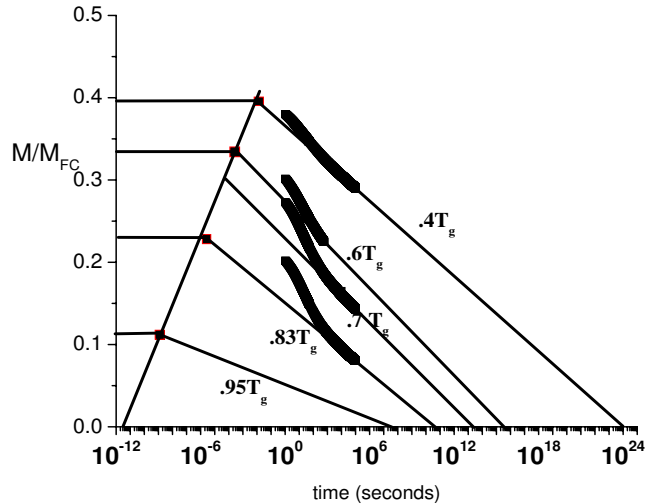}}
\caption[Fit of ZTRM curves]{\label{fig:MuTr} Fit of normalized ZTRM 
curves. The horizontal lines correspond to the value of $M_o$ determined 
from Figure 1. The time on the x-axis begins after the magnetic field is cutoff. The .6 Tg data we did not have a $10^5$s ZTRM curve. A reasonable logarithmic extrapolation from a shorter time measurement was obtained by observing the systematic trend in the slope of the logarithm from a nearby $10^5$s ZTRM curve at .7 Tg. The straight line fit to the .95 Tg data is from Ref. 6(data lost). TRM decays were not measured at .7 Tg so no $M_o$ value is implied.
}
\end{figure}

We continue with a natural extension of this analysis. In experimental ZTRM or TRM experiments, the cooling protocol imposes its own aging.  Therefore the experiments never start with a pure non-aged distribution of states. It is well-known from Field-Cooled - Zero Field-Cooled magnetization experiments that the remnant behavior is very large at low temperatures and decreases to zero at the transition temperature.  It is within this remanence that aging behavior is observed.  A TRM experiment begins by cooling the sample through the transition temperature along the Field Cooled curve and then, after waiting a time 
$t_{\rm w}$, the field is dropped to zero. There is a rapid decrease of the magnetization due to the stationary magnetization decay.  The stationary term is large and rapid with most of the decay occurring well before the measuring time begins in TRM measurements. The magnitude of the stationary term can be observed in Figure 1. Before the TRM decay begins, there is an approximately 60$\%$ drop in the FC magnetization at .4Tg extending up to an approximately 90$\%$ drop in the FC magnetization at .95Tg.  If the stationary decay were instantaneous then the magnetization would drop from $M_{FC}$ to some value $M_o$ after which aging would then proceed.  One can get a reasonably good estimate of $M_o$ from the TRM data displayed in Figure 1. The $t_{\rm w} = 10^4$s waiting time curve is quite flat in the short time regime. There is a slight slope in this time regime but it is likely that a  portion of this is due to the effects of the stationary term. A reasonably good estimate of $M_o$   can therefore be found from a magnetization value just slightly greater than the shortest time value of the $10^4$ second TRM decay. This is the estimated value of the magnetization at which aging begins. The question can then be asked 'What is the time scale at which aging begins.'   While experimentally we cannot produce an idealized starting state after the quench, a uniform distribution within the Barrier Model offers a direction.  If we began with a uniform distribution and looked at the decay with $t_{\rm w}=0s$ we would find that the decay would be logarithmic starting at a time associated with the smallest barrier.  The time for hopping over this smallest barrier is effectively the fluctuation time of the system at any particular temperature.  In Figure 3 we extrapolate the logarithmic term through short time scales to the estimated value of the initial magnetization $M_o$. This was done for all temperatures at which we had a reasonable value of $M_o$. The fluctuation time at each temperature 
is read out as the intersection point of the corresponding curve.   Remarkably these   fluctuation times can be well fit by a straight line on a log-linear plot, i.e. by an Arrhenius function
\begin{equation}
\tau=\tau_{o} \exp{E_a \over {k_bT}}~~.
\end{equation}

The extrapolation of the fit line (Figure 2) to zero magnetization produces the fluctuation time scale at the transition temperature. The value obtained of $\tau_{o}\approx$ 3x$10^{-12}$s corresponds to time scales that might be expected for atomic fluctuations. A rapidly changing fluctuation time scale, as a function of temperature, coupled with the continuum of timescales associated with the barrier space could explain the glassy dynamics associated with the spin glass phase\cite{Binder86}. 
Within this simple model, we are arguing that each domain has a large distribution of time scales (represented by a hierarchical barrier structure) and heterogeneity, at a particular temperature, is determined by the initial starting time scale for a particular domain. The onset of the phase transition is implied by ergodicity breaking and therefore the use of the Barrier Model to describe aging dynamics. In Fig. 2 it can be observed that $\tau_{\rm max}$ (as well as $\tau_o$) decrease as the transition temperature is approached. It will be interesting to observe whether difference between the maximum time and the fluctuation time vanishes at the transition temperature (i.e. $\tau_{\rm max}-\tau_{o}\rightarrow 0s$ as $T \rightarrow T_g$). If this occurs, then as the temperature is lowered through the transition temperature, the spin glass state would appear to evolve (in temporal range) continuously out of the paramagnetic phase. This type of continuous and subtle transition would help explain the lack of any observed discontinuity in the specific heat, at the transition temperature. At this stage, this type of transition is only conjecture and clearly in need of further experimentation.

\section{Simulations}
\label{sec:SclTemp3}

 In this section we perform Barrier Model simulations using experimentally determined parameters.  In general, the Barrier Model has four parameters. The first parameter 
$\alpha$  determines the linear growth rate of the barriers  as a function of Hamming distance and is effectively the size of the smallest barrier.  This parameter therfore also
defines a minimum or fluctuation time scale $\tau_o$. The second parameter N is the number of significant barriers in the system where $N\alpha$ is the size of the largest barrier and defines $\tau_{\rm max}$. The third parameter r is related to the branching ratio and the final parameter is the value of the initial magnetization, $M_o$. For more details of the Barrier Model please see ref\cite{Joh96}. There is a large region of $\alpha$ and r over which simulated TRM experiments exhibit behavior similar to real systems.   For this study we have chosen values of $\alpha$=.4 and r= 1.15. These values were picked somewhat arbitrarily from the range of values which exhibit proper diffusion behavior but were mainly chosen because diffusion on the time scale probed occurs over several hundred barriers. Once r and $\alpha$ were chosen, N was chosen so that $\tau_{\rm max}$ coincided with the maximum time found from the extrapolation of the experimental logrithmic term to zero magnetization. This also has the effect of setting the fluctuation time scale $\tau_o$. Finally, the integrated value of the uniform barrier distribution was set to $M_o$.
Figure 3 displays the simulated TRM plots with the experimental parameters.
 
\begin{figure}
\resizebox{4.0in}{3.0in}{\includegraphics{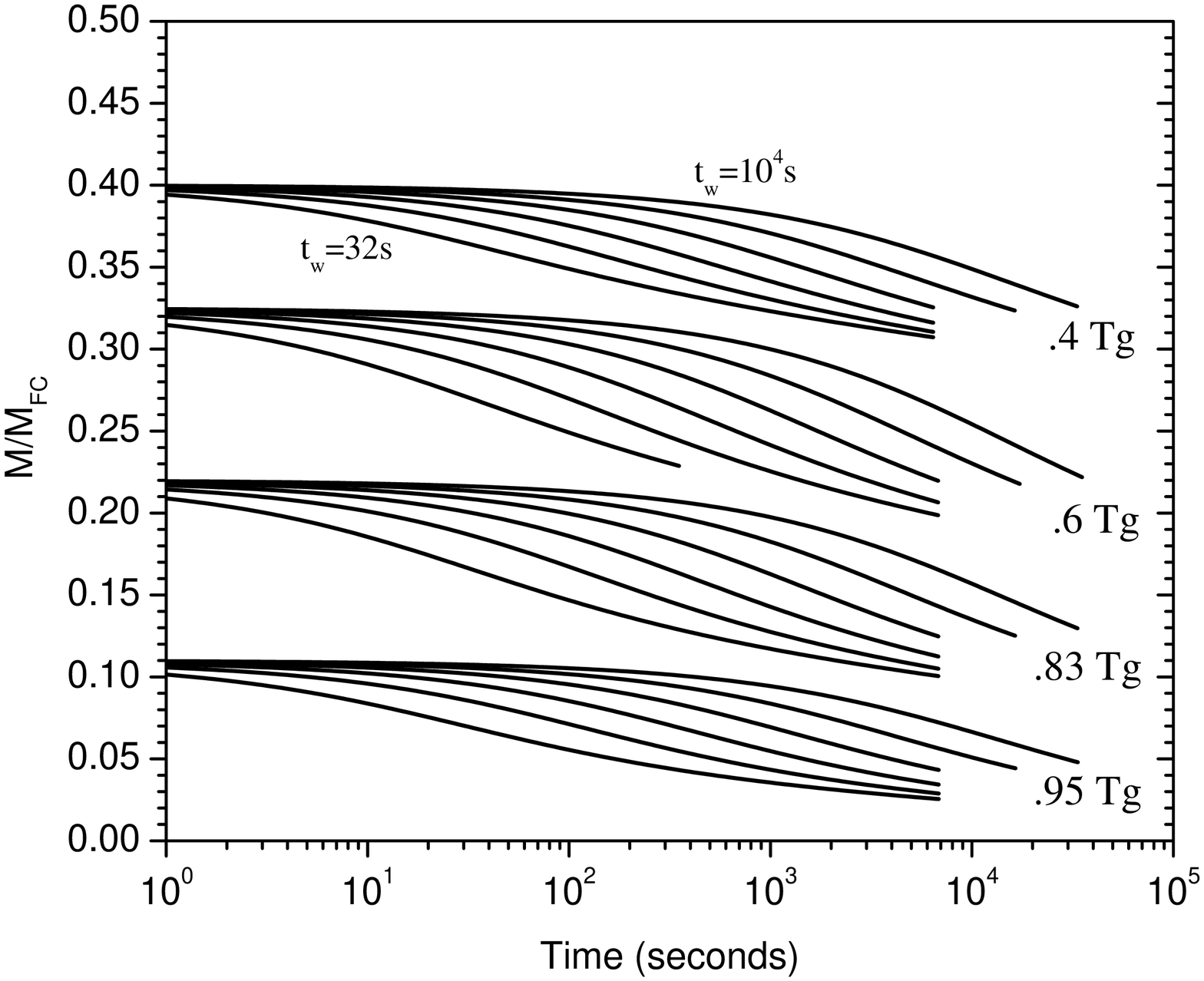}}
\caption[Barrier Model Simulations of TRM decay at Four Temperatures]{\label{fig:MuSclTemp3} Barrier Model simulation decays TRM decays of four different
temperatures .4Tg, .6Tg, .83Tg and .95Tg using experimental parameters. Each set of decay curves at given temperature display simulation TRM decays for waiting times of tw= 32s, 100s, 320s, 1000s, 3200s and 10000s. In each series the 32s curve is the bottom curve and the magnetization increases systematically for increasing waiting time to the uppermost 10000s curve. The time on the x-axis begins after the magnetic field is cutoff.}
\end{figure}

We can see from Figure 3 that the Barrier Model gives a reasonably accurate representation of the experimental data.  There are some significant differences however and some important similarities. One important difference between Figure 3 and Figure 1 is that the experimental data in the short time regime has a slight slope to it whereas the simulations results, especially for the large waiting times, are fairly horizontal.  Some of this difference may be due to the effect of the stationary term on the experimental data at short time scales.  A second important difference is that the range of simulated TRM decays appears to be slightly broader, for each temperature, than the range of the corresponding experimental data.  There are also some interesting similarities.  To begin with the extreme temperatures .4Tg and .95 Tg. appear less broad over the range of waiting times then the intermediate temperatures .83 Tg and .6 Tg.  The similarities are intriguing and we extend our analysis to include $\mu$ scaling of the simulation data to probe these similarities.  Scaling analyses for both the simulated and experimental data are presented in Figure 4.

\begin{figure}
\resizebox{4.0in}{3.0in}{\includegraphics{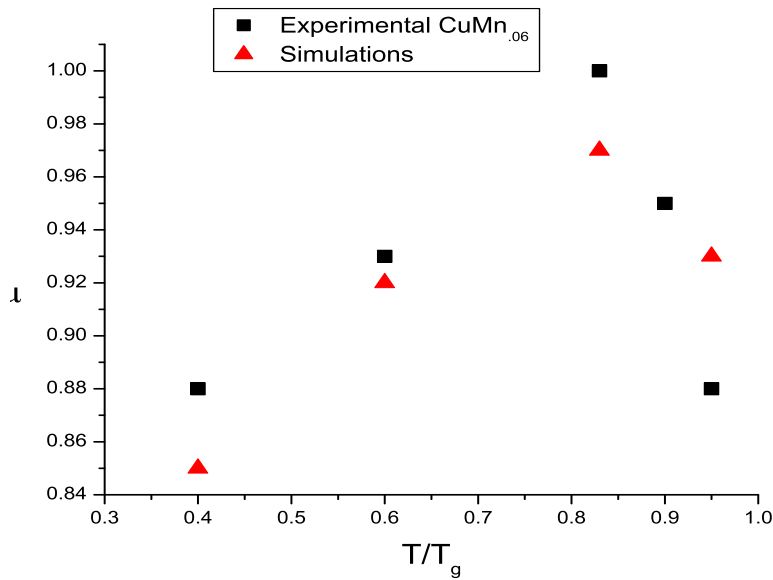}}
\caption[Best fit $\mu$ values for Various
Temperatures]{\label{fig:MuTr2} Best fit $\mu$ values for various
temperatures.  The same eight $t_{w}$'s are used.  The applied
field is still 20G. The measuring temperature is reached with the
fast cooling time protocols.}
\end{figure}

The experimental values of $\mu$, using the fast cooling protocol,  follow a similar pattern to those previously found in the original analyses of scaling behavior in spin glasses\cite{Ocio85, Alba86}. The $\mu$ values are however slightly higher at all temperatures. This result has previously been explained by Parker et al.\cite{Parker06}. It is likely that the smaller $\mu $ values observed for the highest and lowest temperatures reflect the apparent narrowing of the decay set at .4Tg and .95 Tg. The $\mu$ values determined from the simulations are found from scaling the data over the same range of the scaling variable as  the experimental data. It is clear that, once the  the waiting time independent regime is reached, any type of scaling breaks down . The simulation data cannot be scaled over over the entire time range. While scaling of the simulated data was somewhat subjective, full aging ($\mu=1$) was ruled out at all temperatures and subaging was observed. Since aging of an initial distribution concentrated on a  single state leads within the Barrier Model to full aging,  we must conclude in agreement with the assumption of Ref.~\cite{Sibani09}, that subaging is a direct consequence of the distribution of states set up during the thermal quench.

\section{\label{sec:level1}Summary: }
In summary, we have measured the temperature dependent TRM decay
 of CuMn(6$\%$) using fast cooling protocols. We find that the fast
  cooling protocols lead to $\mu$ scaling values larger than previously
   measured values. We have also reanalyzed the ZTRM data taken previously,
    at different temperatures, in conjunction with the TRM data reported in
     this paper. Extracting an initial starting magnetization $M_o$, from 
     the TRM data, and extrapolating the apparent logarithmic long time 
     decay back to this initial magnetization allows us to determine a
      minimum fluctuation time scale at any particular temperature. 
      A plot of this time scale vs. temperature produces apparent 
      Arrhenius like behavior. Extrapolation of this time scale to 
      $M_o=0s$ corresponds to the fluctuation time scale at the 
      transition temperature and appears to be approaching atomic
       fluctuation time scales. The conjecture that cooling the 
       spin glass through the transition temperature leads to a
        heterogeneous distribution of fluctuation time scales
	 provides an explanation for both the long time "logarithmic"
	  decay and subaging.

\medskip

We would like to express great thanks to Ray Orbach for his support and useful discussions. We would also like to thank J.M. Hammann and M. Henkel for useful discussions.

\pagebreak

\end{document}